\documentclass[aps,prl,reprint,superscriptaddress]{revtex4-1}
\usepackage{graphicx}
\usepackage{graphics}
\usepackage{epsfig}
\usepackage{epstopdf} 

\begin{document}


\title{Surface aligned magnetic moments and hysteresis of an endohedral single-molecule magnet on a metal}

\author{Rasmus Westerstr\"om}
\affiliation{Physik-Institut, Universit\"at Z\"urich, Winterthurerstrasse 190, CH-8057 Z\"urich, Switzerland}
\affiliation{Department of Physics and Astronomy, Uppsala University, Box 516, S-751 20 Uppsala, Sweden}
\author{Anne-Christine Uldry}
\affiliation{Swiss Light Source, Paul Scherrer Institut, CH-5232 Villigen PSI, Switzerland}
\author{Roland Stania}
\affiliation{Physik-Institut, Universit\"at Z\"urich, Winterthurerstrasse 190, CH-8057 Z\"urich, Switzerland}
\affiliation{Swiss Light Source, Paul Scherrer Institut, CH-5232 Villigen PSI, Switzerland}
\author{Jan Dreiser}
\affiliation{Institute of Condensed Matter Physics, Ecole Polytechnique F\'{e}d\'{e}rale de Lausanne, CH-1015 Lausanne, Switzerland}
\affiliation{Swiss Light Source, Paul Scherrer Institut, CH-5232 Villigen PSI, Switzerland}
\author{Cinthia Piamonteze}
\author{Matthias Muntwiler}
\affiliation{Swiss Light Source, Paul Scherrer Institut, CH-5232 Villigen PSI, Switzerland}
\author{Fumihiko Matsui}
\affiliation{Physik-Institut, Universit\"at Z\"urich, Winterthurerstrasse 190, CH-8057 Z\"urich, Switzerland}
\affiliation{Nara Institute of Science and Technology (NAIST), 8916-5 Takayama, Ikoma, Nara 630-0192, Japan}
\author{Stefano Rusponi}
\author{Harald Brune}
\affiliation{Institute of Condensed Matter Physics, Ecole Polytechnique F\'{e}d\'{e}rale de Lausanne, CH-1015 Lausanne, Switzerland}
\author{Shangfeng Yang}
\affiliation{Hefei National Laboratory for Physical Sciences at Microscale,
Department of Materials Science and Engineering,
University of Science and Technology of China,
96 Jinzhai Road, Hefei 230026, China}
\author{Alexey Popov}
\affiliation{Leibniz Institute of Solid State and Materials Research, Dresden, D-01069 Dresden, Germany}

\author{Bernd B\"uchner}
\affiliation{Leibniz Institute of Solid State and Materials Research, Dresden, D-01069 Dresden, Germany}
\author{Bernard Delley}
\affiliation{Swiss Light Source, Paul Scherrer Institut, CH-5232 Villigen PSI, Switzerland}
\author{Thomas Greber}
\email{greber@physik.uzh.ch}
\affiliation{Physik-Institut, Universit\"at Z\"urich, Winterthurerstrasse 190, CH-8057 Z\"urich, Switzerland}


\date{\today}

\begin{abstract}
The interaction between the endohedral unit in the single-molecule magnet Dy$_2$ScN@C$_{80}$ and a rhodium (111) substrate leads to alignment of the Dy 4$f$ orbitals. 
The resulting orientation of the Dy$_2$ScN plane parallel to the surface is inferred from comparison of the angular anisotropy of x-ray absorption spectra and multiplet calculations in the corresponding ligand field.
The x-ray magnetic circular dichroism (XMCD) is also angle dependent and signals strong magnetocrystalline anisotropy. 
This directly relates geometric and magnetic structure. Element specific magnetization curves from different coverages exhibit hysteresis at a sample temperature of $\sim4$ K. From the measured hysteresis curves we estimate the zero field remanence life-time during x-ray exposure of a sub-monolayer to be about 30 seconds.
\end{abstract}

\pacs{xxx}
\maketitle

The hollow interior of the fullerene \cite{kroto} carbon cage can be used to encapsulate paramagnetic systems consisting of single atoms, as well as small clusters of different composition \cite{pop13}. A fascinating example is the dysprosium-scandium based endofullerene-series Dy$_n$Sc$_{3-n}$N@C$_{80}~(n=1,2,3)$ where the different stoichiometries result in distinct ground-state properties like tunnelling of magnetization $(n=1)$, remanence $(n=2)$, or frustration $(n=3)$ \cite{westerstromJACS,westerstromPrb14,vieru2013,cimpoesu2014}. 
The strong ligand field, mainly due to the central N$^{3-}$ ion, imposes orientation of the $4f$ shell and therefore non-collinear magnetism. 
In the case of the di-dysprosium compound $(n=2)$, exchange and dipolar coupling between the two magnetic moments stabilizes hysteresis and a large remanence with a relaxation time of one hour at 2 K was found \cite{westerstromPrb14}.
These endofullerenes are thus single-molecule magnets (SMMs) \cite{sessoliNat93,GatteschiBook06,habib2013,woodruff2013,Zhang2013}, a class of magnetic compounds with potential for application in spintronics, quantum computing, and high density storage \cite{leuenbergerNat01,bogani08mNat}. 
Single-molecule magnets have been studied extensively in the bulk phase for the last two decades, but little is known regarding possible modifications to their intrinsic magnetic properties as the molecules are deposited onto substrates or integrated into different device architectures. 
This gap in knowledge can largely be attributed to the fragility of most compounds that have restricted research to a few families of molecules.  
A first proof of principle that molecular nanomagnets can retain their magnetic bistability on a surface was demonstrated for a monolayer (ML) of a Fe$_4$ SMM on a gold surface \cite{ManniniNatMat09}. 
At sub-Kelvin temperatures the Fe$_4$ SMMs exhibited hysteresis, out-of-plane anisotropy, and quantum tunnelling of magnetization (QTM) \cite{ManniniNatMat09,ManniniNature10}. The success of these pioneering experiments is predominantly due to the chemical modification of the Fe$_4$ complex enabling chemically and structurally stable MLs which were prepared \textit{ex situ} from a solvent under ambient conditions. 

Depositing SMMs onto a reactive metal surface such as ferromagnetic substrates  \cite{rizziniPrl,dreiser2014} requires \textit{in situ} preparation under ultra-high vacuum (UHV). 
In this context the mononuclear double-decker complex TbPc$_2$ \cite{Ishikawa03}  is the most studied compound. It has been demonstrated that the magnetic anisotropy is preserved at sub-ML coverage on Cu(100) \cite{stepanow2010} and that the magnetic moments couple antiferromagnetically to thin nickel films on Cu(100) and Ag(100) \cite{rizziniPrl}. 
Magnetic hysteresis comparable to the bulk phase has been observed in thick molecular films \cite{Margheriti10}. Only recently hysteresis with weak remanence was detected in the monolayer regime with graphite \cite{Dalton} and Si \cite{maniniNature14} as a substrate, though the influence of the substrate on the magnetic properties is still poorly understood. 

Endofullerenes that are synthesised with the Kr\"atschmer-Huffmann method are thermally very stable and can be sublimated onto surfaces under UHV. 
For magnetic endohedral units the carbon cage acts as a "spin-shuttle" that protects the spins from chemical interactions.
The robustness of the cages also facilitates imaging and manipulation by scanning probes \cite{butcher03,CPL12,yasutake}.
If the cages have a high symmetry, different orientations of the endohedral units are possible. This decreases the average magnetic moment of a system with more than one molecule, and strategies that circumvent this issue are needed for macroscopic spin alignment.
On metallic surfaces it was shown that also endohedral units may order \cite{treierPRB}, which opens a door for achieving non-vanishing macrospin in zero field.
For the case of Gd$_3$N@C$_{80}$ on Cu(100), a spin system without 4$f$ charge anisotropy, direction dependent magnetic susceptibility was observed, though it could not be directly related to the geometry of the endohedral cluster \cite{hermansPRL}. 

Here we present an x-ray absorption spectroscopy (XAS) study of  Dy$_2$ScN@C$_{80}$ on Rh(111). 
In the first layer the endohedral units and their magnetic moments align with the metal substrate, which is inferred from the angle dependence in XAS and x-ray magnetic circular dichroism (XMCD).  
Hysteresis curves demonstrate that the proximity of the metal surface has a pronounced influence on the magnetic bi-stability. 
Compared to thicker films, which are representative for the bulk phase, the smaller opening of the hysteresis for the sub-ML indicates faster magnetic relaxation times. 
  
\begin{figure}[t]
\begin{center}
\includegraphics[width=8.5cm]{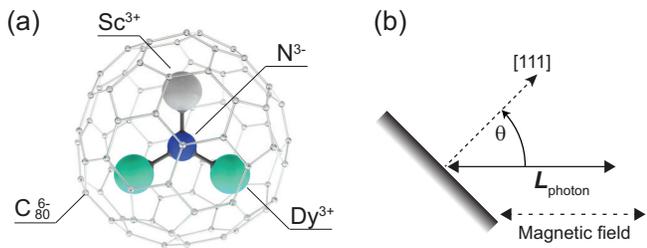}
	\caption{(a) Ball-and-stick model of Dy$_2$ScN@C$_{80}$. (b) Measurement geometry  with the angular momentum of the x-rays $L_{ph}$, parallel or antiparallel to the magnetic field and at an angle of $\theta$ with respect to the normal of the Rh(111) surface.}
\label{fig1}.
\end{center}
\end{figure}

The molecules (see Fig. \ref{fig1}(a)) have been sublimated \textit{in situ} onto the clean Rh(111) substrate, following the recipe in Ref. \cite{treierPRB}.
The sample was cooled in zero magnetic field and the layer thickness is estimated from the x-ray absorption of Dy.
The XAS measurements were performed at the X-Treme beamline \cite{piamontezeXtreme} of the Swiss Light Source. Absorption spectra were acquired by recording the total electron yield in the on-the-fly mode \cite{krempasky}, at sample temperatures of $\sim$4~K and with an external magnetic field applied along the x-ray beam. 

\begin{figure}[t]
\begin{center}
\includegraphics[width=8cm]{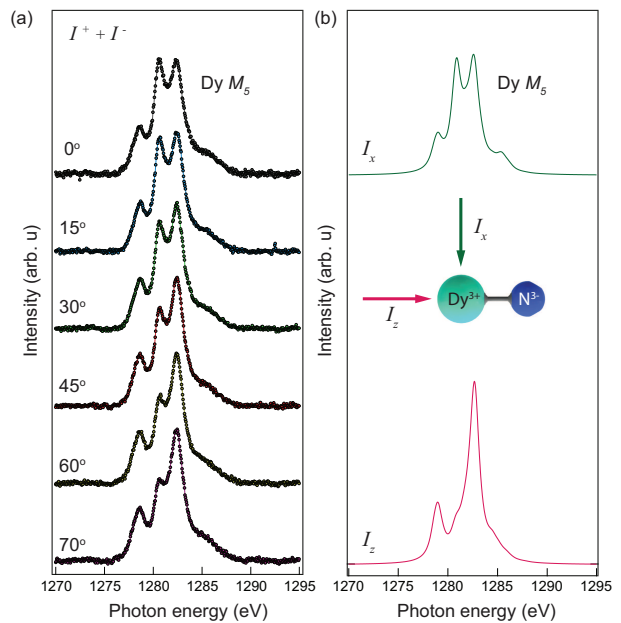}
\caption{(a) XAS measured at the Dy $M_{5}$-edge from a sub-ML of Dy$_2$ScN@C$_{80}$/Rh(111) $T=4$ K, $\mu_0H=6.5$ T, measurement geometry in Fig. \ref{fig1} (b). Each data set is normalized to the integrated intensity. (b) Calculated absorption with the x-ray beam and external field oriented parallel $I_z$, and perpendicular $I_x$, to the magnetic easy-axis (Dy-N bond). }
\label{xas}.
\end{center}
\end{figure}

Fig. \ref{xas} (a) shows x-ray absorption for a sub-ML coverage of Dy$_2$ScN@C$_{80}$ on Rh(111) as a function of the angle $\theta$ between surface normal and x-ray beam (see Fig. \ref{fig1}(b)). The data were recorded over the Dy $M_5$-edge $(3d_{5/2} \rightarrow 4f)$ using right $(I^+)$ and left $(I^-)$ circular polarized x-rays. After background subtraction the XAS  $(I^+ + I^-)$ were normalized to the integrated $M_{5}$ absorption signal in order to compensate for the angular dependence of the total electron yield (TEY).
A significant change in shape of the XAS multiplet spectra is observed as the sample is rotated from normal  $\theta = 0^{\circ}$ to a larger angle of incidence $\theta = 70^{\circ}$. 
This  anisotropy in the XAS is attributed to an anisotropic distribution of the $4f$ electron charges due to their interaction \cite{grootBook} with the ligand field of the central nitrogen ion, the neighbor rare earth ions, and the C$_{80}$ cage. 
This effect would not be present for an isotropic distribution of the endohedral Dy$_2$ScN-units and, therefore, indicates a preferred orientation of the  Dy $4f$ orbitals with respect to the surface.

The orientation of the Dy $4f$ orbitals may be inferred from comparison of the experiments with multiplet calculations. The crystal- or ligand-field multiplet theory for the circularly polarized x-ray absorption is a continuation in a long history of conceptually fairly similar \cite{comment}, close to first principles, calculations with semi-empirical parameters to fine tune the fit to experiment. 
This circumvents unfeasible calculations comprising the coupling to a huge number of electron states of lesser importance for the appearance of the spectrum. 
The endohedral Dy ions are trivalent (Dy$^{3+}$), which leads to a 4$f^9$ groundstate configuration. The final state in the present absorption spectra is consequently $3d^9$  $4f^{10}$.
The ligand field determines the easy axis with a two fold degenerate ground state $\pm J_z$. 
The magnetic field lifts this degeneracy, and induces dichroism \cite{Uldry}.  For low magnetic fields we find a very small influence on the XAS. 
The ligand field, here from a point charge model of the [Dy$_2^{3+}$Sc$^{3+}$N$^{3-}]^{6+}$ ion, describes the site symmetry.
Fig. \ref{xas} (b) displays calculated XAS spectra from the Dy $M_{5}$-edge  with the x-ray beam and an external magnetic field of 6.5 T applied parallel ($I_z$) and perpendicular ($I_x$) to the Dy-N bond. The resemblance of the calculated $I_x$ spectrum and data measured at normal incidence $(\theta = 0^{\circ})$ thus indicates that the endohedral units adopt an orientation parallel to the surface.

\begin{figure}[t]
\begin{center}
\includegraphics[width=8.5cm]{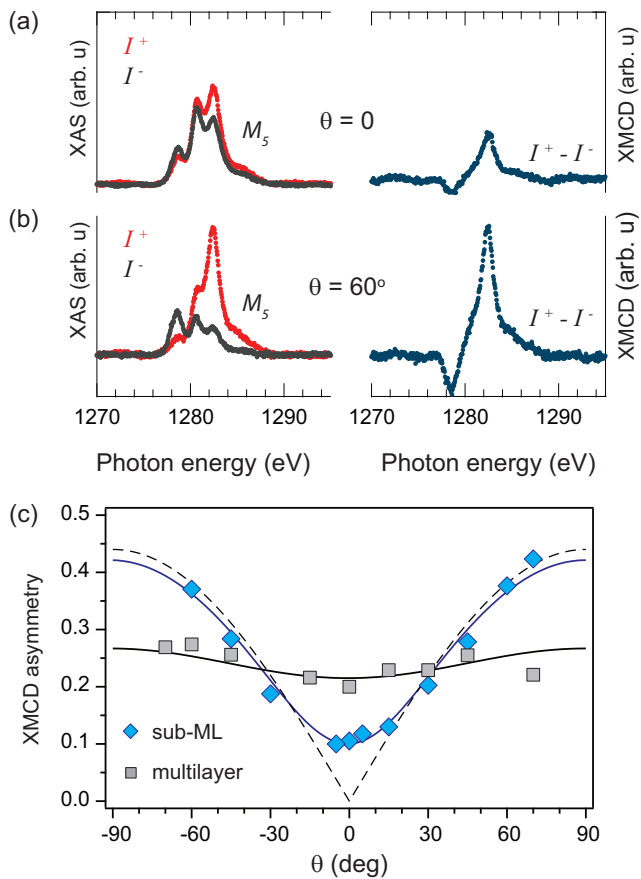}
\caption{Sub-ML of Dy$_2$ScN@C$_{80}$/Rh(111), $T=4$ K, $\mu_0H=6.5$ T, measurement geometry of Fig. \ref{fig1} (b). The polarization dependent XAS spectra (left panel), and the corresponding XMCD spectra (right panel), were measured  at an incidence angle of $\theta = 0^{\circ}$ (a) and $\theta = 60^{\circ}$ (b). (c) Angle dependence of the integrated XMCD signal normalized to the integrated XAS  over the Dy $M_5$-edge. The dashed line corresponds to the expected angle dependence for magnetic moments oriented parallel to the surface, whereas the blue line takes into account a Gaussian distribution, centred in the surface plane and with a standard deviation of $16^{\circ}$.}
\label{xmcd}.
\end{center}
\end{figure}


The magnetism of the system is governed by the spin and orbital moment of the Dy $4f$ electrons that have a total  magnetic moment of $10~\mu_B$ per Dy$^{3+}$ ion \cite{westerstromJACS,westerstromPrb14}. 
Any anisotropy in the spin and orbital moments will give rise to a polarization dependent absorption at the Dy $M_{5}$-edge and an XMCD spectrum $(I^+ - I^-)_{M_5}$. The magnitude of the XMCD signal is determined by the projection of the corresponding magnetic moment $\vec{\mu}_i$ of the absorbing dysprosium ion $i$ onto the direction of the impinging x-rays $\vec{k}$ \cite{stöhrBook}
\begin{equation}
I_{\mathrm{XMCD}} \propto \vec{\mu}_i \cdot \vec{k}
\label{Ixmcd}
\end{equation}  
For an isotropic system  where the magnetic moments are either randomly distributed, or aligned to the external magnetic field,  the resulting XMCD signal in the present measurement geometry is independent of the incidence angle. Any macroscopic magnetic anisotropy is reflected in different XMCD spectra  as a function of  incidence angle. Polarization dependent XAS and corresponding XMCD spectra are shown in Fig. \ref{xmcd} (a) and (b) for incidence angles $\theta$ of $0^{\circ}$ and $60^{\circ}$, respectively. 
Comparison of the two spectra reveals a significant angle dependence which indicates a macroscopic magnetic anisotropy in the sub-ML. Electrostatic interaction with the surrounding ligands, in particular the central N$^{3-}$ ion, results in a strong axial anisotropy which restricts the individual Dy moments $\vec{ \mu}_i$ to orient parallel, or anti-parallel, to the corresponding magnetic easy-axis directed along the Dy$_i$-N bonds \cite{wolfJMag05,westerstromJACS, westerstromPrb14, vieru2013, cimpoesu2014}. The observed magnetic ordering is thus directly related to the strong axial anisotropy of the individual Dy ions {\it{and}} the preferred adsorption geometry of the {\it{endohedral}} cluster, which must be imposed by the surface.

The magnetic anisotropy is quantified in Fig. \ref{xmcd} (c), where XMCD angular dependence is shown. This confirms that the dysprosium moments are predominantly oriented parallel to the surface. A small out-of-plane fraction is inferred from the non-vanishing dichroism at $\theta = 0^{\circ}$. The observed behaviour can be modelled by assuming a Gaussian distribution of the magnetic moments centered in the surface plane ($\theta = 90^{\circ}$). The fit yields a distribution of $90 \pm 16^{\circ}$. This implies that the  Dy-N bonds are not completely parallel to the surface, which is in line with a resonant x-ray photoelectron diffraction study performed on a ML of Dy$_3$N@C$_{80}$ on Cu(111), where the room temperature data indicated a coexistence of planar endohedral units inclined to the surface, and slightly pyramidal configurations parallel to the surface \cite{treierPRB}. 

The square symbols in Fig. \ref{xmcd} (c) correspond to the same measurements performed on a multilayer containing 7 times more molecules. The weak angular anisotropy observed is attributed to the residual influence of the surface.

\begin{figure}[t]
\begin{center}
\includegraphics[width=7.5cm]{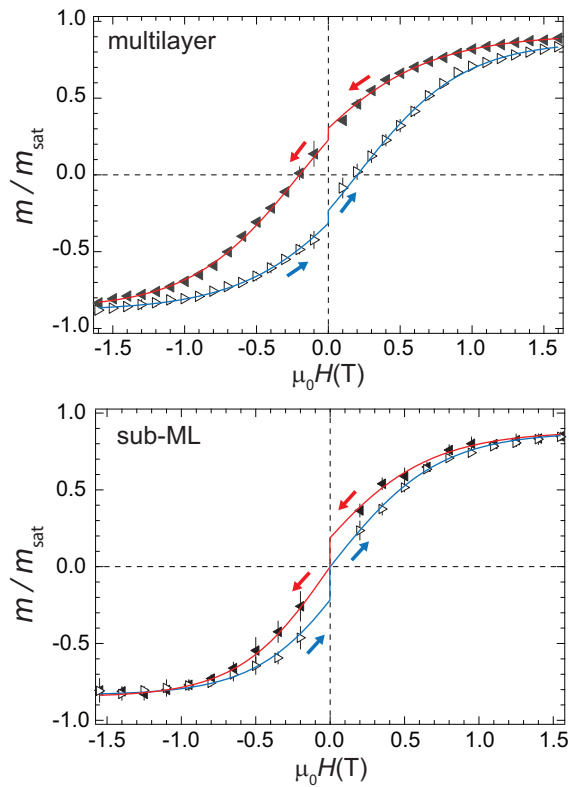}
	\caption{Hysteresis curves measured from a multilayer and a sub-ML of Dy$_2$ScN@C$_{80}$/Rh(111) at a magnetic field sweep rate of 2 T/min and a sample temperature of $\sim4$ K. The x-ray flux was $5 \times 10^{10}$photons/mm$^2$/s. The data were recorded with the x-ray beam and the magnetic field at an incidence angle of $\theta = 60^{\circ}$. The magnetization curves correspond to the average of several independent measurements, where the error bars are the standard deviation at each external magnetic field. The arrows indicate the ramping direction of the magnetic field, the lines are guides to the eye, and $m_{sat}$ is the saturated value at $\pm6.5$ T. The drop in magnetization at zero field is a consequence of the time of 30~s it takes for the magnet to switch polarity. }
\label{hyst}.
\end{center}
\end{figure}

The relaxation time of the magnetic moments in a given environment is a key property of single-molecule magnets. Relaxation times that are slow compared to the time scale of the measurement will result in magnetic hysteresis. 
Fig. \ref{hyst} shows element specific magnetization curves from the Dy $M_5$-edge at a field sweep rate of 2 T/min with the x-rays and the magnetic field at an angle of $60^{\circ}$ with respect to the surface normal. A significant hysteresis is observed for both systems demonstrating that the corresponding relaxation times are slow compared to the measurement time. 
However, comparing the magnetization curves from the two systems indicates that the magnetic bi-stability of Dy$_2$ScN@C$_{80}$ is modified by the proximity of the rhodium metal surface.

Single-ion $4f$ magnets, such as e.g. TbPc$_2$, exhibit poor remanence due to the rapid decay of the magnetization at low fields through QTM. In contrast, for bulk samples of Dy$_2$ScN@C$_{80}$ an exchange and dipole barrier of $0.96\pm 0.1$ meV  suppresses QTM that in turn leads to a significant remanence and coercive field \cite{westerstromPrb14}. This is clearly observed in the multilayer system, where the drop in magnetization at zero field is attributed to the delay of 30 s when changing the polarity of the magnet. From the  25\% decrease in magnetization during these 30 s, we derive a remanence relaxation time of 110 s in the multilayer system.  Compared to bulk samples in the dark \cite{westerstromPrb14} this is about four times  faster, and mainly related to x-ray induced demagnetization \cite{Dreiser14}. 

The remanence-time of the sub-monolayer system is still shorter because the magnetisation vanishes during the switching of the magnet. From the comparison of the two magnetization curves, we can estimate the  remanence time for the sub-ML. Here we assume that the ratio of the hysteresis openings, recorded for a fixed temperature and field sweep rate,  is a relative measure of the relaxation times in the two systems. Under this assumption, we obtain a four times faster relaxation rate in the sub-ML and a remanence time under x-ray irradiation and 4 K of about 30 s. 

The shorter remanence time of the sub-ML may be related to residual interaction of the Rh Fermi sea across the C$_{80}$ shell. 
The substrate interaction that orients the endohedral clusters and spin fluctuations in the metal might impose demagnetizing noise. Furthermore, since the Dy magnetic dipoles lie in a plane their interaction is stronger, which may also accelerate demagnetisation. 
Also, at sub-ML coverage the total electron yield below the $M_5$ edge is about 10\% higher than in the multilayer case and the reduced bi-stability could therefore be demagnetisation due to secondary electrons from the substrate \cite{Dreiser14}. 
However, the opening of the hysteresis demonstrates that the rate at which the magnetization relaxes to its equilibrium value is still slow compared to the measurement time. 

In summary, angle dependent XAS from a sub-ML of Dy$_2$ScN@C$_{80}$ on Rh(111) reveals a one-to-one correspondence between structural and magnetic ordering: The combined effect of the local magnetic easy-axis for the encapsulated Dy ions, {\it{and}} the preferred absorption geometry of the endohedral cluster, indictes surface aligned $4f$ moments and a macroscopic non-collinear anisotropy.  At a sample temperature of $\sim4$~K we observe a hysteresis in the sub-ML.
Although orientational ordering of endohedral molecules at surfaces, as well as magnetic hysteresis of bulk samples of such molecules have been shown, we demonstrate here the orientational structural and magnetic ordering of surface adsorbed endohedral molecules creating a stable macrospin for molecular sub-monolayers.
This bi-stablity is observed at a one order of magnitude higher sample temperature than previously reported for the $3d$-based Fe$_4$ SMMs \cite{ManniniNatMat09,ManniniNature10} and thus paves the way for many more experiments, like scanning tunneling spectroscopy on single molecules.  

We gratefully acknowledge financial support from the Swiss National Science Foundation (SNF project 200021 129861, 147143, and PZ00P2-142474), the Swedish research council (350-2012-295) and the Deutsche Forschungsgemeinschaft (DFG project PO 1602/1-1). 


\begin{thebibliography}{99}
\bibitem{comment}
Contrary to earlier approaches, our method treats S-O; electron-electron  and crystal field at the same level in a single variation.
\end{thebibliography}

\end{document}